\documentclass[aps,pra,twocolumn,floats,showpacs]{revtex4}

\usepackage{graphicx}
\usepackage{amsmath}
\usepackage{amssymb}
\usepackage[usenames,dvipsnames]{color}
\usepackage{textcomp}
\usepackage{xfrac}
\usepackage{bm}
\usepackage[normalem]{ulem}

\newcommand{\Trev}{{T_\text{rev}}}

\newcommand{\E}{{\cal E}}

\newcommand{\D}{{\Delta}}

\newcommand{\beq}{\begin{equation}}
\newcommand{\eeq}{\end{equation}}
\newcommand{\bea}{\begin{eqnarray}}
\newcommand{\eea}{\end{eqnarray}}

\begin{document}
\title{Experimental demonstration of coherent control in quantum chaotic systems}

\date{\today}
\author{M.~Bitter and V.~Milner}
\affiliation{Department of  Physics \& Astronomy and The
Laboratory for Advanced Spectroscopy and Imaging Research
(LASIR), The University of British Columbia, Vancouver, Canada \\}

\begin{abstract}{

We experimentally demonstrate coherent control of a quantum system, whose dynamics is chaotic in the classical limit. Interaction of diatomic molecules with a periodic sequence of ultrashort laser pulses leads to the dynamical localization of the molecular angular momentum, a characteristic feature of the chaotic quantum kicked rotor. By changing the phases of the rotational states in the initially prepared coherent wave packet, we control the rotational distribution of the final localized state and its total energy. We demonstrate the anticipated sensitivity of control to the exact parameters of the kicking field, as well as its disappearance in the classical regime of excitation.
}
\end{abstract}


\maketitle


Control of molecular dynamics with external fields is a long-standing goal of physics and chemistry research. Great progress has been made by exploiting the coherent nature of light-matter interaction. At the heart of coherent control is the interference of quantum pathways leading to the desired target state from a well-defined initial state \cite{ShapiroBrumerBook}. In this context, an exponential sensitivity to the initial conditions, characteristic for classically chaotic systems, poses an important question about the controllability in the quantum limit (for a comprehensive review of this topic, see \cite{Gong2005}). As the underlying classical ro-vibrational dynamics of the majority of large polyatomic molecules is often chaotic, the answer to this question has far reaching implications for the ultimate prospects of using coherence to control chemical reactions.

Success in steering the outcome of chemical reactions by means of feedback-based adaptive algorithms \cite{Assion1998}, using the methods of optimal control theory \cite{Judson1992}, proved that such control is feasible. Theoretical works on quantum controllability in the presence of chaos, both in general \cite{Rice2000} and with regard to specific molecular systems \cite{Gong2001JCP, Abrashkevich2002}, pointed at the importance of coherent evolution. To investigate the roles of coherence, chaoticity and quantumness further, Gong and Brumer considered a paradigm system for studying quantum effects on classically chaotic dynamics - the quantum kicked rotor (QKR) \cite{Gong2001PRL, Gong2001JCP, Gong2005}. The latter is known to exhibit dynamical localization (closely related to Anderson localization in disordered solids \cite{Anderson1958, Fishman1982}), in which quantum interferences suppress the classically chaotic diffusion after the ``quantum break time'' \cite{Casati1979, Izrailev1980}. Gong and Brumer demonstrated that the energy of the localized state can be controlled by modifying the initial wave packet. They showed that quantum coherences, as opposed to the classical structures in the rotor's phase space \cite{Shapiro2007}, are indeed responsible for the achieved control over the chaotic dynamics of the QKR.

\begin{figure}
\centering
 \includegraphics[width=1.0\columnwidth]{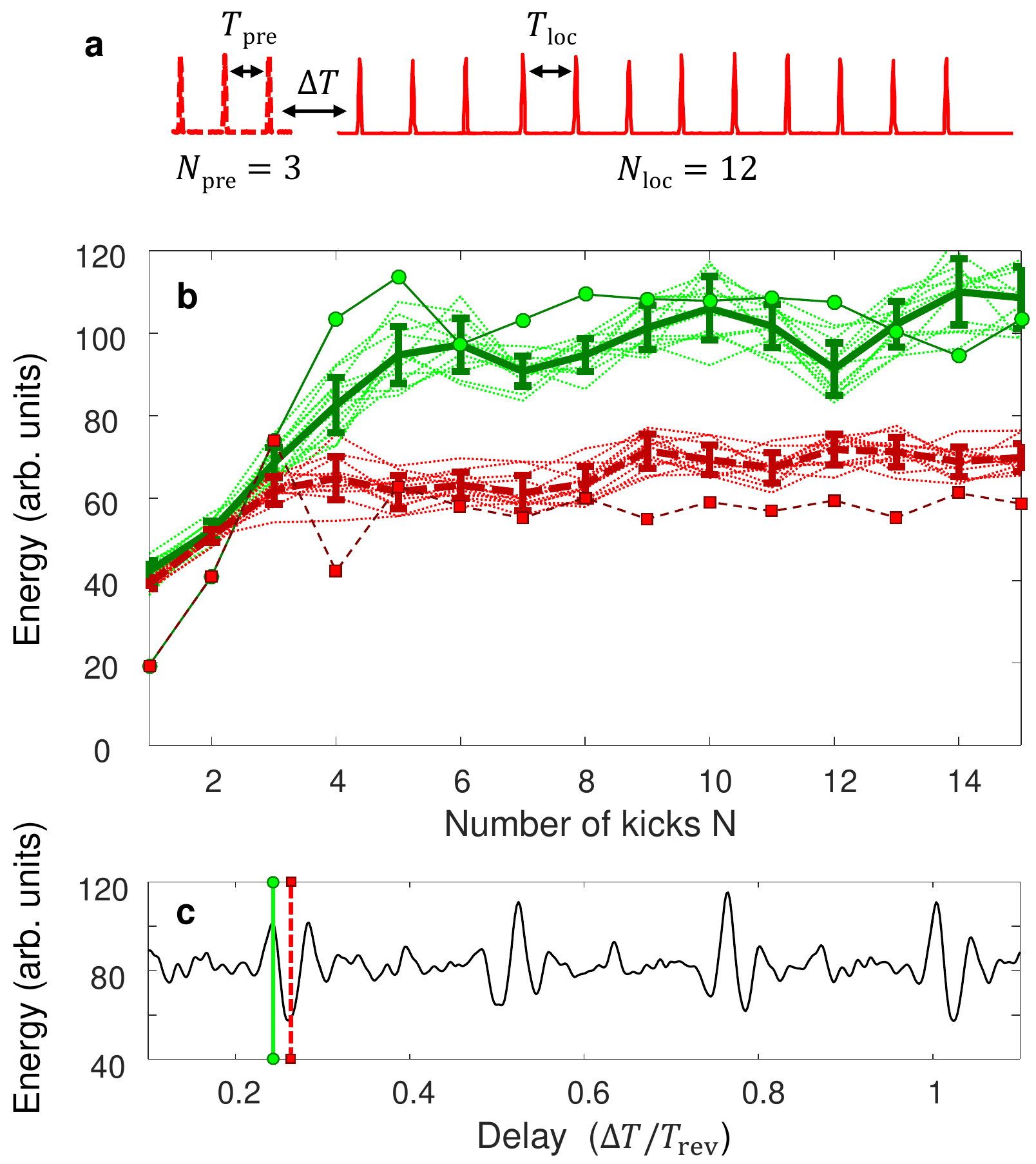}
     \caption{(color online) (\textbf{a}) Train of fifteen laser pulses, used in this work, with three variable time constants indicated by horizontal arrows. (\textbf{b}) Rotational energy of oxygen molecules as a function of the number of kicks $N$. Shown are thirteen experimental realizations (dotted lines) for each of the two control scenarios corresponding to a maximum (upper green lines, at $\D T_1=0.243\ \Trev$) and a minimum in rotational energy (lower red lines, at $\D T_2=0.264\ \Trev$). The corresponding average values are plotted as the green solid line and the red dashed line, respectively, with error bars representing one standard deviation. In comparison, the numerical calculations are indicated by connected green circles ($\D T_1$) and red squares ($\D T_2$). (\textbf{c}) Numerically calculated dependence of the final rotational energy on the delay $\D T$. Two vertical lines mark the experimental delays $\D T_1$ (solid green) and $\D T_2$ (dashed red).}
  \label{Fig:Energy}
\end{figure}

In this report, we present an experimental proof of the Gong-Brumer control scheme. Following a theoretical proposal of Averbukh and co-workers \cite{Floss2012, Floss2013}, we investigate the dynamics of true quantum rotors by exposing diatomic molecules to a periodic sequence of ultra-short laser pulses. A number of representative QKR effects have already been studied in laser-kicked molecules \cite{Cryan2009, Zhdanovich2012, Floss2015a, Kamalov2015, Bitter2016b}, including our recent observation of the formation of localized rotational states under periodic kicking \cite{Bitter2016c}. Here, we prepare the molecules in a coherent rotational wave packet and control the localization process by varying the relative phases of the initial states. The preparation is executed by preceding the long localizing pulse sequence (12~pulses) with a shorter sequence of 3~pulses tuned to a fractional quantum resonance [Fig.~\ref{Fig:Energy}(\textbf{a})]. The time delay $\Delta T$ between the two pulse trains, and hence the relative quantum phases of the initial states, serves as a ``control knob'' defining the amount of the rotational energy, absorbed before its further growth is suppressed by localization.

The interaction of a diatomic molecule with a periodic train of $N$ linearly polarized laser pulses, not resonant with any electronic transition, is described by the following Hamiltonian:
\begin{equation}\label{Eq-Hamiltonian}
    \hat{H}=\frac{\hat{J}^2}{2I} - \hbar P \cos^2(\theta) \sum_{n=0}^{N-1} \delta(t-nT),
\end{equation}
where $\theta$ is the angle between the molecular axis and the laser polarization axis, $\hat{J}$ is the angular momentum operator, $I$ is the molecular moment of inertia, $T$ is the train period and $\hbar$ is the reduced Planck constant. The laser-induced rotational dynamics of a molecule is determined by the effective Planck constant $\tau=\hbar T/I$ and a kick strength $P= \D\alpha /(4\hbar) \int \E^2(t) dt$, where $\D\alpha$ is the molecular polarizability anisotropy and $\E(t)$ is the temporal envelope of the pulse. In the classical limit, the dynamics is governed by a single stochasticity parameter $K=\tau P$. For $K\gtrsim1$, which applies to all of our experimental conditions, the underlying classical dynamics is predominantly chaotic and exhibits unbounded diffusive energy growth \cite{Izrailev1990}.

The discreteness of the rotational spectrum of the QKR results in quantum resonances whenever $\tau =2\pi p/q$, where $p$ and $q$ are integers \cite{Izrailev1980, Wimberger2003, Floss2012}. Equivalently, this condition can be expressed as $T/\Trev=p/q$, with $\Trev=2\pi I/\hbar$ being the so-called revival period. Tuning the train period to match a quantum resonance enables an efficient excitation of multiple rotational states with growing (from kick to kick) rotational energy. On the other hand, away from quantum resonances, dynamical localization suppresses the rotational energy growth after the quantum break time. In this work, we employ the resonant driving of the quantum kicked rotors to control their further localization by a non-resonant pulse train.


A sequence of 15 laser pulses, shown in Fig.~\ref{Fig:Energy}(\textbf{a}), is generated in an optical system, shown schematically in supplementary Fig.~\ref{Fig:Setup}(\textbf{a}) and described in detail in \cite{Bitter2016a}. A Ti:Sapphire femtosecond laser system produces pulses of 130~fs full width at half maximum (FWHM) at a central wavelength of 800~nm. The pulse sequence is created in a standard `$4f$' pulse shaper \cite{Weiner2000} and is further amplified in a multi-pass amplifier to reach a kick strength of $P=3.8$ ($\sim 10^{13}$~W/cm$^{2}$ at 10~Hz repetition rate). It consists of two independent parts. First three ``preparation'' pulses are separated in time by $T_\text{pre}=0.237\ \Trev$, close to a fractional quantum resonance at $T=\sfrac{1}{4}\ \Trev$, and are used to excite a broad rotational wave packet \cite{Bitter2016b}. The period $T_\text{loc}$ of the second ``localizing'' train of 12 pulses is chosen between $0.26\ \Trev$ and $0.27\ \Trev$, corresponding to the effective Planck constant of $1.6<\tau<1.7$. This window is chosen so as to avoid strong fractional quantum resonances of low orders. The corresponding range of the stochasticity parameter $6.2<K<6.5$ lies deep in the classically chaotic regime. The time delay between the two pulse sequences is scanned around the quarter revival time, between $\Delta T/\Trev=0.223$ and 0.284, where we anticipate the highest degree of control, as discussed below.

The excitation light is focused in a gas of oxygen molecules, rotationally cooled down to 25~K in a supersonic expansion. Coherent molecular rotation modulates the refractive index of the gas and results in the appearance of Raman sidebands in the spectrum of a weak narrow-band probe pulse [for details see references \cite{Bitter2016b, Bitter2016c} and supplementary Figure \ref{Fig:Setup}(\textbf{b,c})]. Each Raman peak is shifted from the central probe frequency by the amount that depends on the rotational quantum number $J$, while its intensity $I_J$ is proportional to the square of the population of the corresponding level $P_J$ \cite{PopulationApprox}.
The latter allows us to determine the rotational energy, absorbed by the molecules, as $\sum_J E_J P_J$, where $E_J=B J (J+1)$ with the rotational constant $B$. To compare the experimental findings with the results of numerical simulations, we solve the Schr{\"o}dinger equation, using the above described Hamiltonian (\ref{Eq-Hamiltonian}).

Our main result is shown in Fig.~\ref{Fig:Energy}(\textbf{b}), where we plot the rotational energy of oxygen molecules, measured after each of 15 laser pulses for a number of pulse trains, all with $T_\text{loc}=0.267\ \Trev$. By design, the first three preparation pulses in all trains lead to a fast growth of molecular energy. When the delay $\Delta T$ to the next twelve pulses is set to $\Delta T_1=0.243\ \Trev$ (upper green lines), the energy growth continues for a few more kicks and ceases after that, reflecting the dynamical localization of the molecular angular momentum \cite{Bitter2016c}. Different thin lines correspond to different experimental runs, with their average indicated by the thick green curve. On the other hand, when the very same localizing pulse sequences are separated from the preparation pulses by $\Delta T_2=0.264\ \Trev$, the suppression of the energy growth occurs much earlier and results in a lower (by $40\pm7\%$) energy of the final localized states (lower red lines).

The results of the equivalent numerical calculations are shown in Fig.~\ref{Fig:Energy}(\textbf{b}) by connected green circles for the delay $\Delta T_1$ and red squares for $\Delta T_2$. Despite the used approximation of infinitely short $\delta $-kicks, the numerical results are in good qualitative agreement with the observations. We further exploit the numerical model for calculating the dependence of the rotational energy on the single control parameter $\Delta T$, plotted in Fig.~\ref{Fig:Energy}(\textbf{c}). The availability of control is apparent around fractional revivals, $\Delta T/\Trev=\sfrac{1}{4}, \sfrac{1}{2},\sfrac{3}{4}$ and 1, which suggests an intuitive picture of its mechanism. The first kick from the localizing pulse train either continues the quantum-resonant excitation of the preparation sequence or opposes it, affecting the energy level, at which the rest of the train localizes the system. The dephasing of the rotational states in the prepared wave packet leads to a loss of control between the fractional revivals. The two vertical lines mark our experimental values of  $\Delta T$ in Fig.~\ref{Fig:Energy}(\textbf{b}).

\begin{figure}
\centering
 \includegraphics[width=1.0\columnwidth]{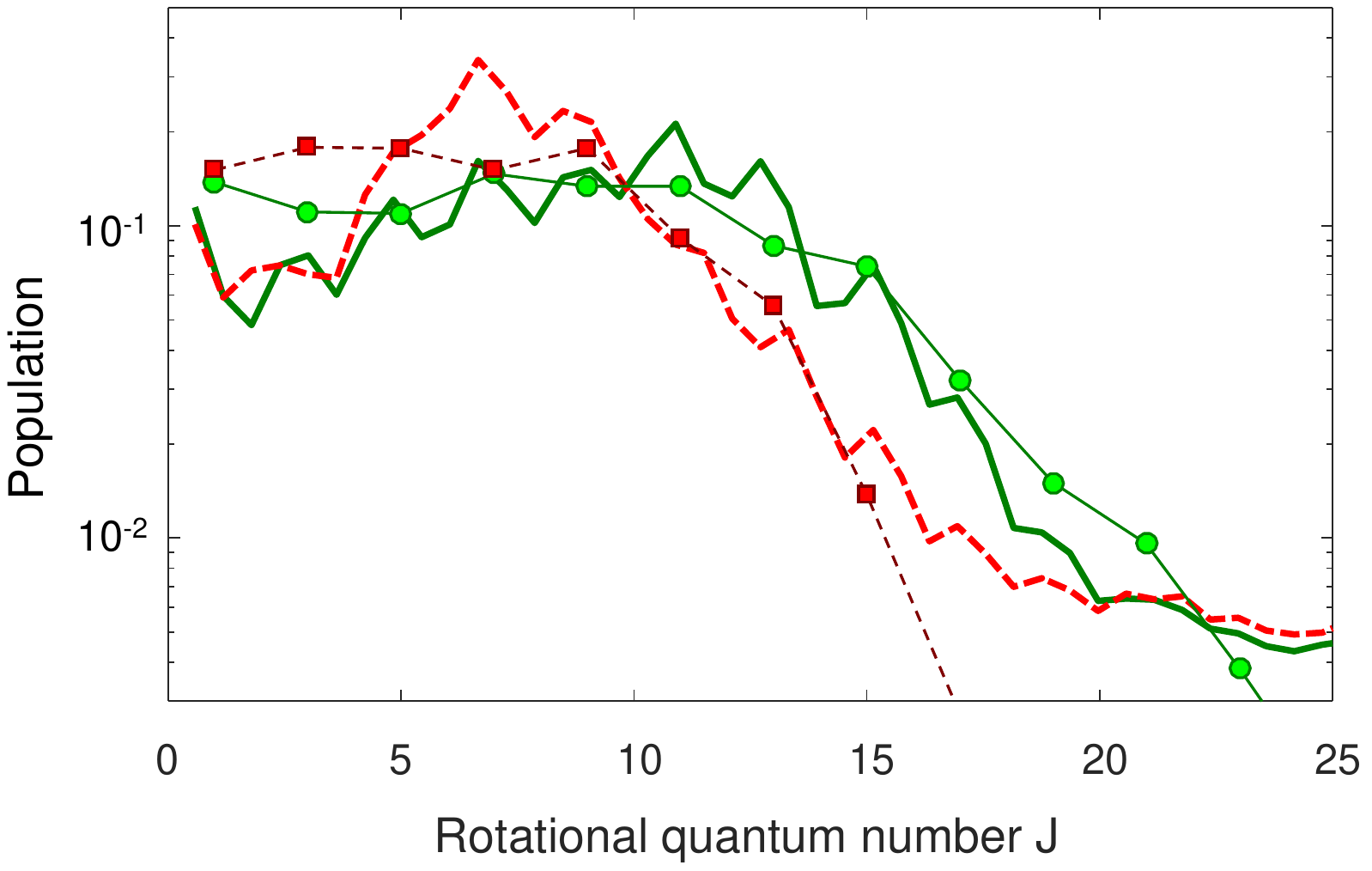}
     \caption{(Color online) Localized population distribution of oxygen molecules excited by a train of 15~pulses with $P=3.8$. The pulse train parameters are given in the text. Plotted is the experimentally retrieved average population distribution (thick lines, no markers) and the numerically calculated one (markers, connected by thin lines). Only odd values of $J$ are allowed due to the nuclear spin statistics of oxygen. The distributions correspond to the high (upper green lines, $\D T=\D T_1$) and low (lower red lines, $\D T=\D T_2$) localization energy in Fig.~\ref{Fig:Energy}.
      }
  \label{Fig:Population}
\end{figure}

The described control mechanism is also evident from the experimentally retrieved  average distributions of the localized angular momentum, shown in Fig.~\ref{Fig:Population} by thick lines with no markers. Solid green and dashed red traces correspond to the localized wave packets with higher and lower rotational energies, respectively. As the higher energy clearly correlates with the broader wave packet, the achieved control can be attributed to populating different sets of quasienergy (Floquet) states \cite{Floss2013}. Because each wave packet contains more than a single quasienergy state, the distributions are not expected to (and, indeed, do not) exhibit exponential line shapes \cite{Gong2001PRL}.

Numerically calculated population distributions, corresponding to the experimental parameters for the high and low energy localized wave packets, are shown in Fig.~\ref{Fig:Population} with connected green circles and red squares, respectively. Despite the approximations in the population retrieval from the measured Raman spectra, the simulated and experimental distributions show qualitative agreement down to the instrumental noise floor around $P_J\approx 5\cdot 10^{-3}$. The systematic underestimation of the experimentally extracted population at low rotational states is due to the neglected dependence on the magnetic quantum number \cite{Bitter2016c} and the effect of spin-rotation coupling in oxygen.

\begin{figure}
\centering
 \includegraphics[width=1.0\columnwidth]{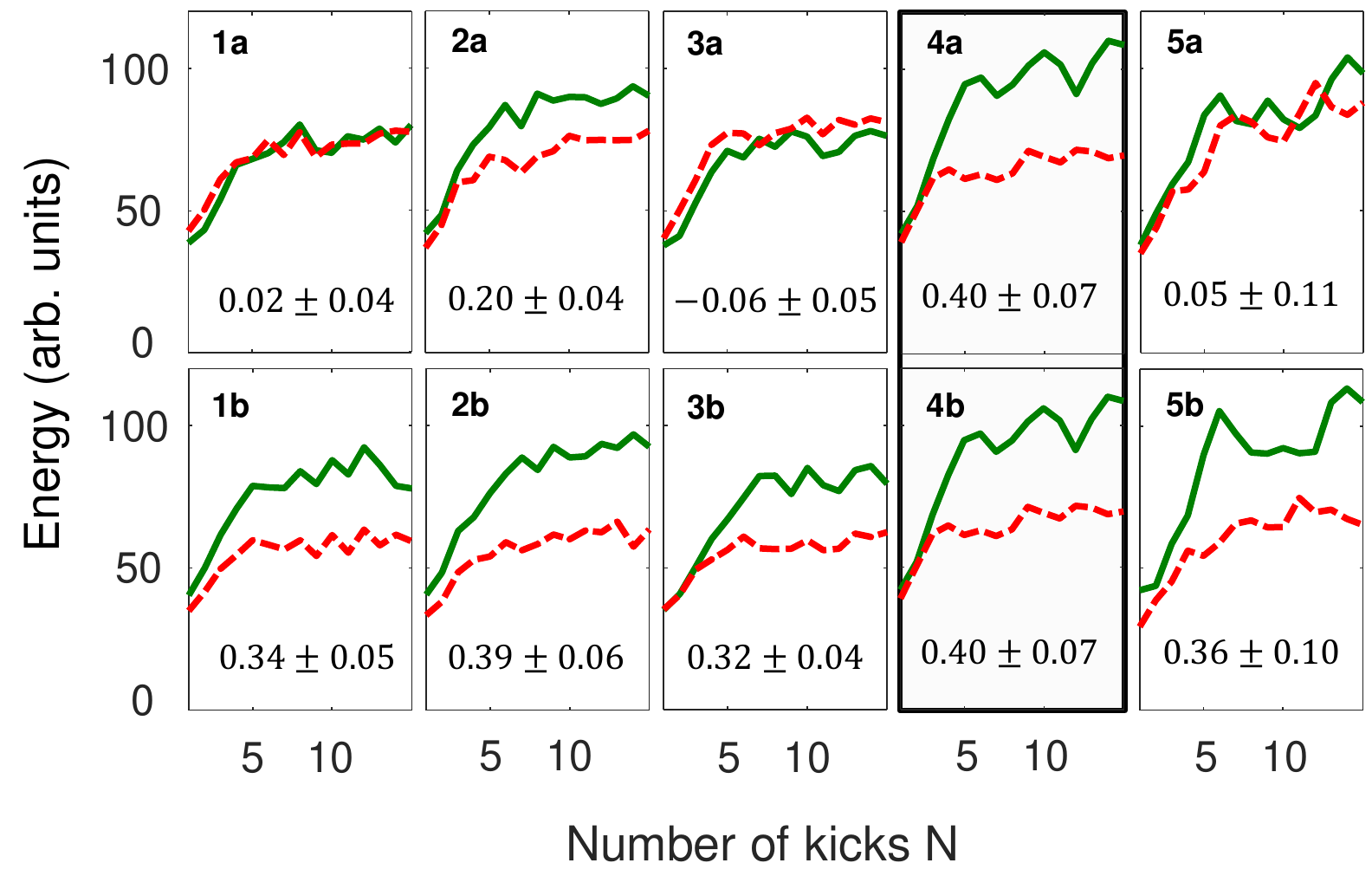}
   \caption{(Color online) Top row (\textbf{a}): rotational energy for both time delays, $\D T_1=0.243\ \Trev$ (sold green line) and $\D T_2=0.264\ \Trev$ (dashed red line)  for a set of five different $T_\text{loc}$ periods  (\textbf{1})-(\textbf{5}). Other parameters of the localizing train remain unchanged. Bottom row (\textbf{b}): for the same five values of $T_\text{loc}$, delays $\D T_1$ and $\D T_2$ are individually adjusted for the respectively highest and lowest energy of the localized state. The degree of control is given in each plot. Column (\textbf{4}) is equivalent to Fig.~\ref{Fig:Energy}.}
  \label{Fig:Sensitivity}
\end{figure}

The stability of the implemented control scheme with respect to the underlying classically chaotic dynamics is analyzed in Fig.~\ref{Fig:Sensitivity}. In the top row (\textbf{a}) we show the dependence of the rotational energy on the period of the localizing train $T_\text{loc}$. As earlier, the value of the control parameter is either $\Delta T_1=0.243\ \Trev$ (sold green line) or  $\Delta T_2=0.264\ \Trev$ (dashed red line). Shown is a representative set for five values of $T_\text{loc}/\Trev$: (\textbf{1}) 0.260, (\textbf{2}) 0.261, (\textbf{3}) 0.263, (\textbf{4}) 0.267 and (\textbf{5}) 0.270. The respective degree of control, defined as $\frac{E_1-E_2}{(E_1+E_2)/2}$ with $E_i$ being the final rotational energy for the delay $\Delta T_i$, is shown at the bottom of each plot. We observe wide fluctuations from a total loss of control in the panels (\textbf{1a,3a,5a}) to the maximum control of about 40\% in panel (\textbf{4a}).

High sensitivity of the QKR dynamics to the exact train period is well expected \cite{Shapiro2007} and can be attributed to the existence of fractional resonances, $T_\text{loc}/\Trev=p/q$, where quantum diffusion is accelerated. Yet despite the observed sensitivity of the control, we found that it can be successfully regained by optimizing the control parameter, i.e. the delay time $\Delta T$, for each individual realization of the localizing train. In the bottom row (\textbf{b}) of Fig.~\ref{Fig:Sensitivity}  we demonstrate this sustained controllability, which supports the assumption of its coherent nature. We note that our numerical calculations of the molecular response to the localizing train of infinitely short $\delta$-kicks (not plotted) show more stable control, which suggests that the finite experimental pulse width may also contribute to the observed sensitivity.

\begin{figure}
\centering
 \includegraphics[width=1.0\columnwidth]{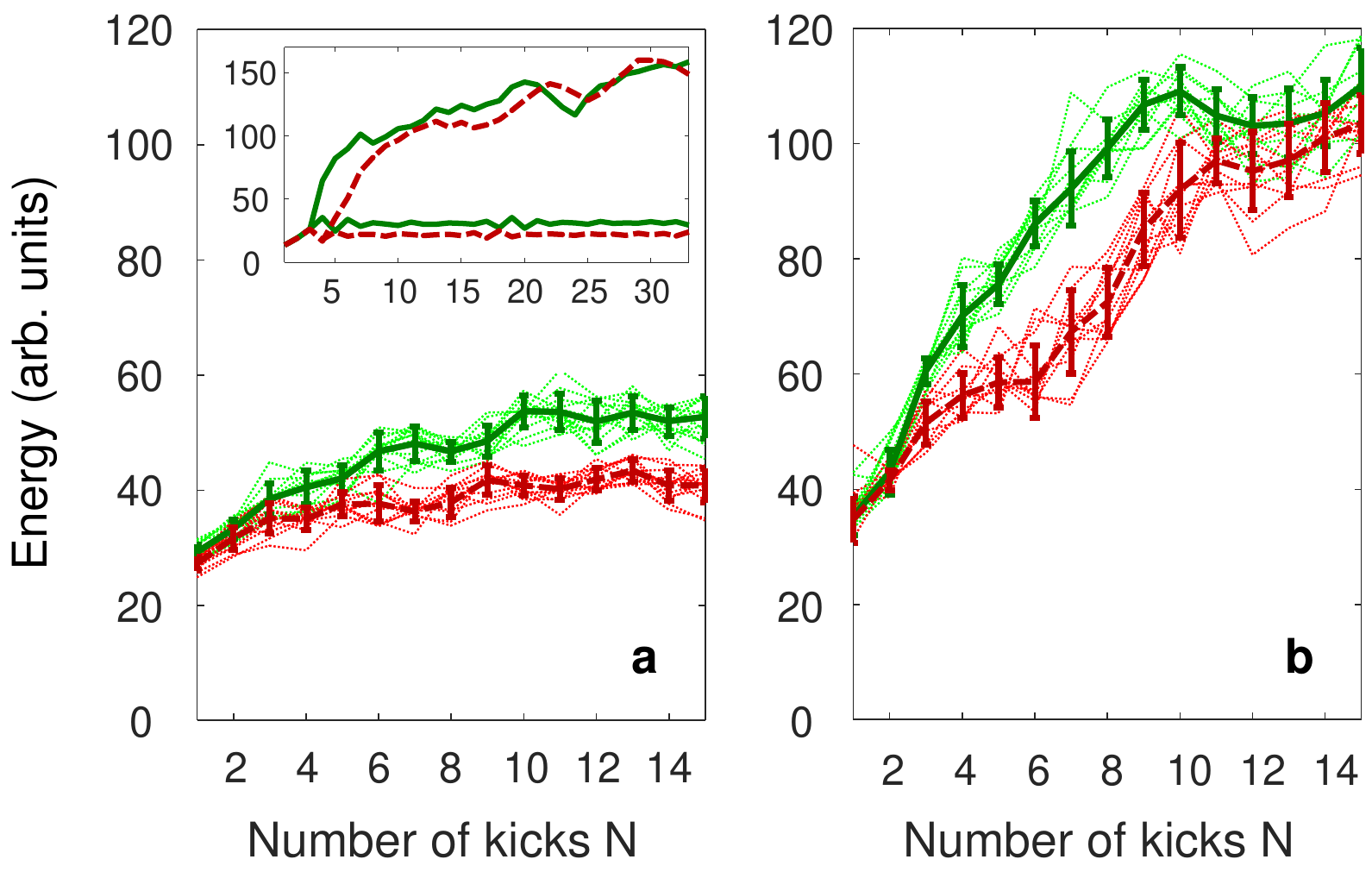}
   \caption{(Color online) Same as Fig.\ref{Fig:Energy}, but for two different values of the effective Planck constant: $\tau=1.7$ (\textbf{a}) and $\tau=0.6$ (\textbf{b}). The stochasticity parameter is held constant at $K=3.4$. The inset gives a numerical comparison of both regimes, with lower (top) and higher (bottom) value of $\tau $, for longer sequences of infinitely short $\delta$-kicks.}
  \label{Fig:QMvsCL}
\end{figure}

To distinguish between the quantum and classical mechanisms of the achieved control, we analyze its dependence on the effective Planck constant $\tau $. Smaller values of $\tau $, realized with shorter periods of the pulse train, take us closer to the classical limit (i.e. the well known standard map \cite{Casati1979}), at which the dynamics is less sensitive to the discreteness of the QKR spectrum. We keep the stochasticity parameter constant at $K=\tau P=3.4$, large enough to stay in the predominantly chaotic regime, and reduce $\tau $ while increasing the kick strength $P$ proportionally. As demonstrated in Fig.~\ref{Fig:QMvsCL}(\textbf{a}), for $\tau =1.7$, the localized states are reached after about 10 kicks. The quantum break time is longer than the one in Fig.~\ref{Fig:Energy}(\textbf{a}) due to the lower kick strength (2 vs. 3.8). The maximum degree of control ($25\pm3\%$) is established between $\Delta T_1=0.232\ \Trev$ (solid green line) and $\Delta T_2=0.263\ \Trev$ (dashed red line).

Figure~\ref{Fig:QMvsCL}(\textbf{b}) shows the result of the same experiment with $\tau =0.6$ and $P=5.6$. Although the dynamics is still sensitive to $\Delta T$, the unbounded growth of rotational energy results in the decreasing relative difference between the two cases and, therefore, diminishing degree of coherent control. The apparent energy saturation at later times is due to the finite duration of our laser pulses. The latter results in the suppressed excitation of rotational states with $J>20$, reached in the absence of localization. An oxygen molecule occupying these states rotates by $\gtrsim 90$\textdegree\ during the length of the pulse, which lowers its effective kick strength and prevents further diffusion. Numerical simulations with a larger number of $\delta$-kicks, shown in the inset, better illustrate the transition between the controlled localization at $\tau =1.7$ (bottom two lines) and the uncontrolled classical diffusion at $\tau =0.6$ (top two lines).  Evidently, the latter effective Planck constant is small enough for the diffusive energy growth to persist.

In summary, we used diatomic molecules exposed to a sequence of strong laser pulses as true quantum kicked rotors, well known for their chaotic dynamics. We demonstrated that despite the exponential loss of memory about the initial conditions in the classical limit, the relative phases in the initial coherent superposition of rotational states can be used to control the QKR dynamics in the absence of noise or decoherence. Adjusting a single control parameter results in the changing rotational distribution of the final localized state: its peak is shifted from a low (here, $J=7$) to a high ($J=11$) angular momentum. This corresponds to a relative change in the rotational energy, absorbed by the laser-kicked molecules. The coherent quantum nature of the control mechanism is evident from the demonstrated high sensitivity of the localized wave packet to the exact period of the pulse train, and the ability to regain control for any value of that parameter. Driving the system closer to the classical limit, while maintaining the same degree of stochasticity, results in a gradual loss of control. Studying chaotic dynamics with molecular rotors may lead to interesting unforeseen effects of centrifugal distortion, external fields or inter-molecular collisions on the controllability of quantum chaotic systems.

We thank Ilya~Averbukh for stimulating discussions and Johannes~Flo{\ss} for his help with numerical calculations. This research has been supported by the grants from CFI, BCKDF and NSERC.


\begin{thebibliography}{25}
\bibitem{PopulationApprox}
The proportionality is exact for a single populated initial state and an exponentially localized final distribution. We verified numerically that it is a reasonable approximation to retrieve the population distribution from our Raman spectrum, even at a non-zero temperature.\\
\end{thebibliography}

\newpage


\section*{Supplementary Material}
\label{SI1}
\begin{figure}[h]
\centering
 \includegraphics[width=0.86\columnwidth]{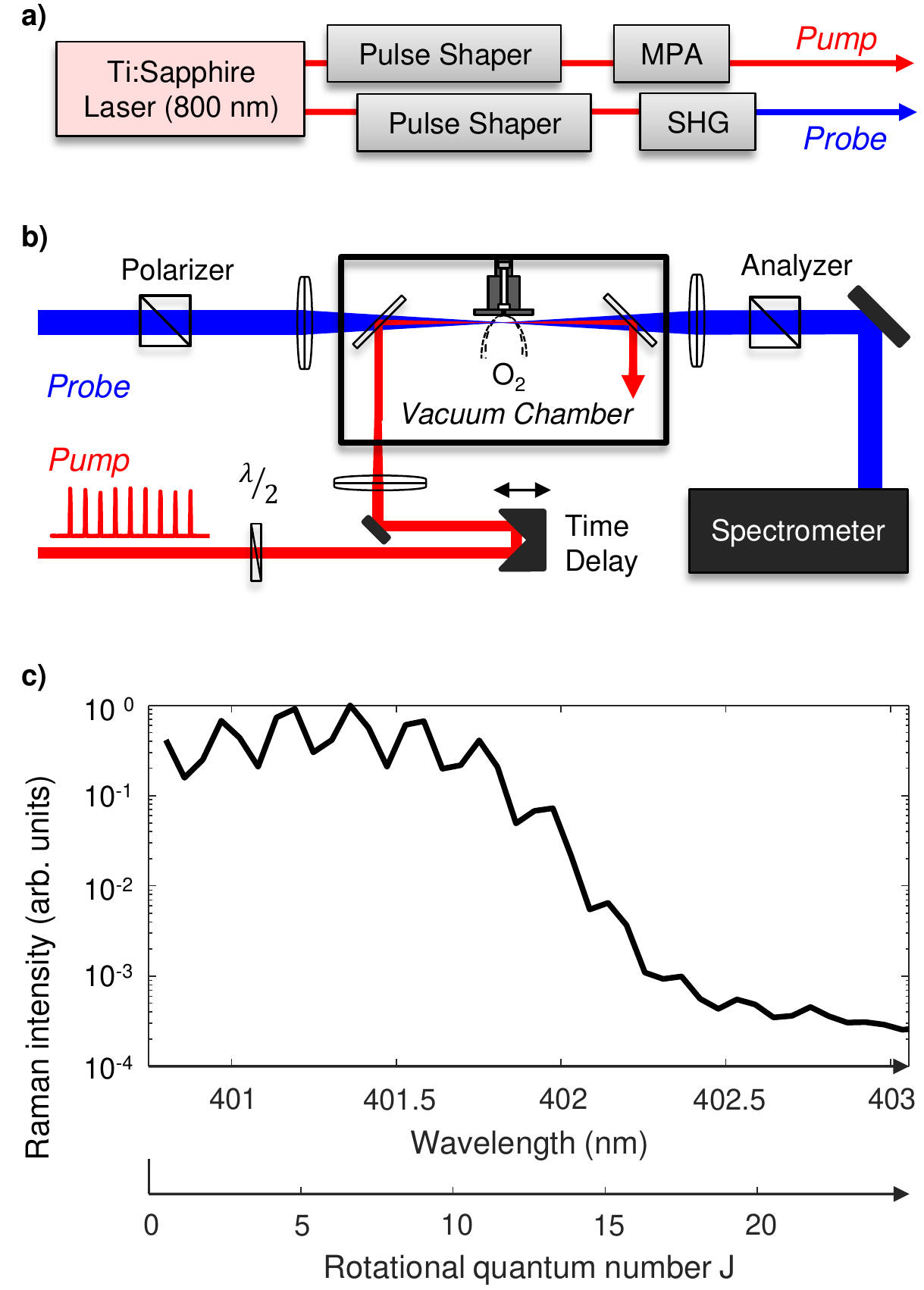}
     \caption{ (\textbf{a}) Diagram of the pump and probe sources originating from a Ti:Sapphire femtosecond laser system with a central wavelength of 800~nm. Long pulse trains are generated via a pulse shaper after which their energy is increased by a multi-pass amplifier (MPA). Another pulse shaper is used to narrow the spectral bandwidth of the probe pulse, whose central wavelength is shifted to $400$~nm by means of second harmonic generation (SHG) in a nonlinear crystal. The probe pulse with a spectral width of 0.15~nm (FWHM) is linearly polarized at 45\textdegree\  with respect to the pulses in the pump train.
      (\textbf{b})
     Scheme of the experimental setup. The train of strong femtosecond pulses and the delayed weak probe pulse are combined on a dichroic beam splitter and focused into a vacuum chamber, where they intersect a supersonic jet of oxygen molecules. The change of probe polarization is analyzed as a function of wavelength by means of two crossed polarizers and a spectrometer. Detrimental effects of spatial averaging are minimized by making the probe beam significantly smaller than the pump (FWHM beam diameters of $20~\mathrm{\mu m}$ and $60~\mathrm{\mu m}$, respectively). We use a $500~\mathrm{\mu m}$ diameter pulsed nozzle, operating at the repetition rate of 10~Hz and the stagnation pressure of 33~bar, to achieve a rotational temperature of about $25$~K at a distance of 2~mm from the nozzle. 
       (\textbf{c}) Typical Raman spectrum of oxygen after the excitation with three preparation pulses (kick strength of $P=3.8$). The measured frequency shift is translated to the rotational quantum number $J$, shown along the lower axis.}
  \label{Fig:Setup}
\end{figure}

\end{document}